\documentclass[]{interact}

\usepackage{dsfont}
\usepackage{xcolor}
\usepackage{tcolorbox}
\usepackage{cancel}
\usepackage{multicol, amsthm}

\usepackage{parskip}
\usepackage{epstopdf}
\usepackage[caption=false]{subfig}

\usepackage[doublespacing]{setspace}
\usepackage[numbers,sort&compress]{natbib}
\bibpunct[, ]{[}{]}{,}{n}{,}{,}

\theoremstyle{plain}

\theoremstyle{definition}

\theoremstyle{remark}

\begin{document}

\articletype{Original Research Paper}

\title{Is it even rainier in \emph{North} Vancouver? A non-parametric rank-based test for semicontinuous longitudinal data}

\author{
\name{Harlan Campbell\textsuperscript{a}\thanks{CONTACT Harlan Campbell. Email: harlan.campbell@stat.ubc.ca}}
\affil{\textsuperscript{a}University of British Columbia Department of Statistics
Vancouver, BC, Canada, V6T 1Z2}
}

\maketitle

\begin{abstract}

When the outcome of interest is semicontinuous and collected longitudinally, efficient testing can be difficult.  Daily rainfall data is an excellent example which we use to illustrate the various challenges.  Even under the simplest scenario, the popular `two-part model', which uses correlated random-effects to account for both the semicontinuous and longitudinal characteristics of the data, often requires prohibitively intensive numerical integration and difficult interpretation.  Reducing data to binary (truncating continuous positive values to equal one), while relatively straightforward, leads to a potentially substantial loss in power. {We propose an alternative: using a non-parametric rank test recently proposed for joint longitudinal survival data.  We investigate the potential benefits of such a test for the analysis of semicontinuous longitudinal data with regards to power and computational feasibility.}

\end{abstract}


\pagebreak

\small{\begin{quote}{\footnotesize{There is booming rain, whispery rain, rain that lulls you to sleep, and rain on the leaves which sings you awake; there is soft rain, hard rain, sideways rain, rain that makes you instantly wet, and rain that leaves soft kisses on your cheek, like the wings of a butterfly.}}\end{quote}
\hfill{\footnotesize{\emph{Vancouver In The Rain, }}}\\
\vspace{-0.75cm}
\hfill {\footnotesize{ Regan D'Andrade, 1999}}}
 \vspace{1cm}

\section{Introduction}

{Semicontinuous} data is characterized by a mixture of zeros and continuously distributed positive values.  Typically, the proportion of observed values equal to zero is substantial and the positive values observed exhibit right-skewness and heteroscedasticity.   What's more, semicontinuous data is distinct by the fact that zeros represent `non-occurrences', rather than simply being the result of truncated or censored negative values.  While this paper only considers data in which the outcome variable is {semicontinuous}, there are related challenges with data in which predictor variables are semicontinuous (also known as `spike at zero' data) \cite{lorenz2017modeling}.

Daily rainfall is an excellent example of semicontinuous data.  On dry days, rainfall is zero,  while on rainy days, rainfall is a positive continuous number of millimetres (mm). A natural approach is to consider semicontinuous data as the result of a two-part process: the `binary part' determining whether an observation is zero (i.e. `Is it raining today? yes/no'), and the `continuous part'  determining the positive value of the observation given that it is nonzero (i.e. `Given that it is raining today, how much?'). In order to test the null hypothesis of no treatment effect (i.e. `Is it rainier over here than over there?'), two-part models were first considered in econometrics \cite{xiao1999comparison, 
 tu1999wald}.

By tracking daily rainfall data, one can test whether or not two places (e.g. the city of Vancouver and the neighbouring city of North Vancouver) are equally rainy, but things are complicated by the unavoidable correlation between yesterday's, today's, and tomorrow's weather.  To address the inherent correlation within \emph{longitudinal} semicontinuous data, Olsen and Schafer (2001) (O\&S) \cite{olsen2001two} 
 proposed a two-part model that accounts for the correlation between a given subject's repeated observations by including subject-specific random effects.  The important idea put forth by 
O\&S \cite{olsen2001two}
 is that one can account for the semicontinuous distribution of the outcome variable by allowing the random-effects included in the binary and the continuous parts of the two-part model to be correlated.  Despite computational challenges, the two-part model for longitudinal semicontinuous data has inspired numerous extensions and variations.

Liu et al. (2012) \cite{liu2012analyzing}
 extended the model by allowing positive values to follow different distributions: (a) a generalized gamma distribution, (b) a log-skew-normal distribution, and (c) a normal distribution after the Box-Cox transformation.   
Tom, Su and Farewell (2016) \cite{tom2016corrected}
 introduced correct formulation for marginal inference.   Most recently, 
Lo (2015) \cite{lo2015assessing}  
 fit four versions: (1) a logit-log-normal random effects model, (2) a two-part logit-truncated normal random effects model, (3) a two-part logit-gamma random effects model, and (4) a two-part logit-skew normal random effects model.

Unfortunately, obtaining maximum likelihood parameter estimates for the two-part model remains difficult.   In the statistical literature, the process has been described as `challenging'
\cite{liu2012analyzing},
 `a unique challenge', 
\cite{olsen1998class}
and one that can `lead to severe computational problems' \cite{su2009bias}.  O\&S discuss several strategies including the Monte Carlo EM algorithm, which they determine to be `far too slow' \cite{olsen2001two}.  Further  details on the EM approach are provided in earlier work 
 with the same conclusion: `it is computationally accurate but too slow for practical use' \cite{olsen1998class}.

Tooze et al. (2002) \cite{tooze2002analysis} 
recommend using adaptive Gaussian quadrature.  This approach is made relatively easy due to implementation available within the SAS PROC NLMIXED procedure.   However, the method is computationally demanding, with 
Su et al. (2009)
noting that: `even with properly standardized explanatory variables and the simplest model with 2 correlated random intercepts, it can take several hours to fit using the SAS NLMIXED procedure' \cite{su2009bias}.  After much consideration, 
O\&S
recommend using Fisher scoring to maximize a sixth-order Laplace-expansion approximation of the likelihood noting that, parameter estimates are obtained relatively quickly under favourable circumstances  (when implemented with a Fortran-90 program) \cite{olsen2001two}.   Under less favourable conditions, the Laplace-expansion strategy appears less capable.   When fitting the two-part model to artificial data simulated to mimic longitudinal 1,000 subject survey data, 
 O\&S 
find that the algorithm fails to converge in over 60\% of runs \cite{olsen2001two}.   Despite this, the authors remain optimistic concluding that: `This leads us to believe that when the algorithm does converge, the estimated coefficients and standard errors are quite reliable.'  And in the event that convergence fails, analysts will: `typically abandon the model and try a simpler one'  \cite{olsen2001two}.
 
What simpler model should analysts turn to in the event that convergence fails?   
 Su et al. (2009) \cite{su2009bias} 
warn against using two independent models (i.e. the two-part model with non-correlated random-effects).  Ignoring the correlation between random effects can introduce bias in the estimation of the regression coefficients and variance components due to an `informative cluster size' problem.  This is due to the fact that parameters in the binary part of the model inevitably impact the number of observations ($m_{i}^{*}$) for the continuous part of the model.

A random-effects Tobit model 
\cite{epstein2003Tobit}, 
may also be problematic \cite{kim2009two}. According to some, the Tobit model lacks the richness required for semicontinuous data and is therefore an ``unattractive'' option \cite{schafer1999modeling}.  The Tobit model assumes that there is both an unobserved normally distributed latent variable, and an observed outcome variable.  The observed outcome variable is equal to the latent variable whenever the latent variable is above a known threshold, and is equal to the threshold otherwise (a result of being censored, or of being bellow the limit of detection); for more details see \cite{twisk2009longitudinal}.   O\&S  note that, in addition to the problematic, potentially ``dubious'' \cite{schafer1999modeling} interpretation, the Tobit model is inappropriate if the semicontinuous outcome is indeed the result of two separate processes, rather than a single process and a censoring mechanism \cite{olsen2001two}.   It remains unclear how this ``inappropriateness'' and ``unattractiveness'' may impact hypothesis testing with regards to validity and efficiency.

A useful alternative could be a non-parametric rank-based test proposed by  Lin et al. (2013) \cite{lin2013estimating} 
for the analysis of combined survival and quantitative outcomes.  Rank tests have the desirable feature that null distributions of test statistics are exact, and do not depend on detailed parametric assumptions about error distributions \cite{brown2005standard}.  This article investigates the merits of such an approach.  After a brief overview of the two-part model, in Section 2, and details concerning the proposed rank-based test in Section 3,  the results of a simulation study comparing a variety of methods across a range of scenarios will be discussed in Section 4.  An application to rainfall data will be presented in Section 5.  Section 6 concludes with recommendations for future research.

\section{The two-part model}

Let us review the most basic implementation of the two-part model.  Consider the simplest scenario in which a semicontinuous outcome variable and a binary covariate of interest (e.g. treatment= drug vs. placebo) are observed for each of $i=1, ..., n$ subjects at each of $j=1,...,m_{i}$ time-points.  The order of the $m_{i}$ time-points may or may not be important and the distribution of the outcome may change (linearly) with time.

Let $Y_{ij}$ be the semicontinuous outcome for individual $i$ at time-point $j$.   Let $X_{i}$ be an $m_{i}$ by 3 matrix, with a first column of 1s (intercept) and a second column taking values of the covariate of interest (e.g. =1 for drug, =0 for placebo) for individual $i$ at time-point $j$.  The third column of $X_{i}$ can be equal to the $m_{i}$ time-points.  Under the {two-part} model, $Y_{ij}$ is recoded as two variables defined as follows:
  \begin{multicols}{2}
\[
    U_{ij}=\left\{
                \begin{array}{ll}
                  1, $ if $ Y_{ij} \ne 0\\
                 0, $ if $  Y_{ij} = 0
                \end{array}
              \right.
\quad \quad \quad \quad \quad \quad \textrm{and:}
 \]

\[ 
\quad
V_{ij}=\left\{
                \begin{array}{ll}
                  g(Y_{ij}), $ if $ Y_{ij} \ne 0\\
                 $\emph{irrelevant},  if $  Y_{ij} = 0
                \end{array}
              \right.
  \]
  \end{multicols}
  
where $g()$ is a monotonic increasing function chosen such that $V_{ij}$ is approximately Gaussian.  As such, for $i$ in 1, ..., $n$,  $U_{i}$ is a binary vector of length $m_{i}$, and $V_{i}$ is a positive valued vector of length $m^{*}_{i} = \sum_{j=1}^{m_{i}} I(Y_{ij} > 0)$, where $I()$ is the standard indicator function.  (If, for a given subject $i$, only zeros are observed, $m^{*}_{i}=0.$)  For $U$, a mixed logistic regression model \cite{wang1998mixed} is defined by:
\begin{eqnarray*}
\eta_{i} = X_{i}\boldsymbol{\beta} +Z_{i}c_{i}, & \quad  \textrm{with:} \quad \boldsymbol{\beta} = (\beta_{0}, \beta_{1}, \beta_{2}),\quad 
\end{eqnarray*}

where $\eta_{i}$ is a $m_{i}$-length vector taking values $\eta_{ij}$ = log $\pi_{ij}/(1- \pi_{ij})$, $\pi_{ij} = Pr(U_{ij} = 1)$.    
The `random intercept model' defines $Z_{i}$ equal to a $m_{i}$-length vector of 1s (\emph{intercept}).  Alternatively, the `random slope model' defines $Z_{i}$ as a $m_{i}$ by 2 matrix with the first column equal to 1s (\emph{intercept}) and the second column equal to the $m_{i}$ time-points (\emph{slope}), with $c_{i}=(c_{i1}, c_{i2})$.  

For $V$, a Gaussian random-effects regression model \cite{laird1982random} is defined by:
\begin{eqnarray*}
V_{i}=  X^{*}_{i}\boldsymbol{\gamma}+ Z^{*}_{i}d_{i}+\epsilon_{i}, 
&\quad \textrm{where:} \quad \boldsymbol{\gamma} = (\gamma_{0}, \gamma_{1}, \gamma_{2}) ,\quad
\end{eqnarray*}

with $\epsilon_{i} \sim N(0, \sigma^{2}I)$ and $X_{i}^{*}$ and $Z_{i}^{*}$ are $m^{*}_{i}$-subsets of $X_{i}$ and $Z_{i}$ for points at which $Y_{ij} > 0$.  Under the `random slope model',  $d_{i}=(d_{i1}, d_{i2})$.  In order to account for the relationship between $U$ and $V$, the two parts of the model are connected by allowing $c_i$ and $d_{i}$ to be correlated:

\begin{equation}
 b_{i} = \left( \begin{array}{c}
c_{i}\\
d_{i} \end{array} \right)    \sim N \left(0, \psi= \left( \begin{array}{c c}
\psi_{cc} \quad \psi_{cd}\\
\psi_{dc} \quad \psi_{dd} \end{array} \right)   \right) \quad \textrm{Random intercept model}
\end{equation}

\begin{equation}
 b_{i} = \left( \begin{array}{c}
c_{i1}\\
c_{i2}\\
d_{i1}\\
d_{i2} \end{array} \right)    \sim N \left(0, \psi= \left( \begin{array}{c c c c}
\psi_{c_{1}c_{1}} \quad \psi_{c_{1}c_{2}} \quad \psi_{c_{1}d_{1}} \quad \psi_{c_{1}d_{2}}\\
\psi_{c_{2}c_{1}} \quad \psi_{c_{2}c_{2}} \quad \psi_{c_{2}d_{1}} \quad \psi_{c_{2}d_{2}}\\
\psi_{d_{1}c_{1}} \quad \psi_{d_{1}c_{2}} \quad \psi_{d_{1}d_{1}} \quad \psi_{d_{1}d_{2}}\\
\psi_{d_{2}c_{1}} \quad \psi_{d_{2}c_{2}} \quad \psi_{d_{2}d_{1}} \quad \psi_{d_{2}d_{2}}\\
 \end{array} \right)   \right) \quad \textrm{Random slope model}
\end{equation}

It is worth noting that even under the very simple scenario described, the number of parameters required for the two-part model is substantial, see Table 1.  Under the two-part model, the null hypothesis of no treatment effect is composite and defined as:  $\textrm{H}_{0}$:  $\beta_{1} = 0$ and $\gamma_{1} = 0$.   One-sided alternatives are not trivial, see \cite{zhou2017multivariate}.
Oftentimes, one is interested in determining whether or not the rate at which the outcome variable changes, is dependant on treatment.  In this situation, a treatment by time interaction must also be included in the model.  This introduces at least two additional parameters to estimate ($\beta_{3}$ and $\gamma_{3}$), and the null hypothesis is even more complex: $\textrm{H}_{0}$:  $\beta_{1} = 0$, $\beta_{3} = 0$,   $\gamma_{1} = 0$,  and $\gamma_{3} = 0$.

\begin{table}
\tbl{Summary of the `fixed effects' + `random effects' parameters required of each model.}{
\begin{tabular}{lcc} \toprule
\textbf{    }& \textbf{Random } & \textbf{Random } \\
\textbf{    }& \textbf{ Intercept} & \textbf{ Slope} \\
Fixed effects    & $\beta_{0}$  , $\beta_{1}$  , $\beta_{2}$    & $\beta_{0}$  , $\beta_{1}$  , $\beta_{2}$ \\
 & $\gamma_{0}$ , $\gamma_{1}$ , $\gamma_{2}$   & $\gamma_{0}$ , $\gamma_{1}$ , $\gamma_{2}$   \\
\hline
Variance  &$\sigma^{2}$ &$\sigma^{2}$ \\
 & $\psi_{cc}$ & $\psi_{c_{1}c_{1}}$, $\psi_{c_{1}c_{2}}$, $
\psi_{c_{1}d_{1}}$, $\psi_{c_{1}d_{2}}$\\
& $\psi_{cd}$& $\psi_{c_{2}c_{2}}$, $\psi_{c_{2}d_{1}}$, $\psi_{c_{2}d_{2}}$\\
 & $\psi_{dd}$ & $\psi_{d_{1}d_{1}}$, $\psi_{d_{1}d_{2}}$, $\psi_{d_{2}d_{2}}$ \\
\hline
Random effects    &$c_{i}$ & $c_{i1}$ ,  $c_{i2}$\\
    &$d_{i}$ & $d_{i1}$ , $d_{i2}$\\
\hline
\textbf{Number of } &   & \\
\textbf{Parameters } &  10 + 2$n$ & 17 + 4$n$\\

\hline
\end{tabular}}
\end{table}

Many more possibilities exist for the two-part model.  The complexity can be substantially greater when the number of covariates increases, different random effects setups are required, and changes over time are more elaborate (e.g. time-dependent covariates, lag-effects, serial correlation).  In their concluding remarks, O\&S note that hierarchical clustering can be quite common (e.g. neighbourhoods nested within cities) \cite{olsen2001two}.  In order to account for this type of correlation structure, a much larger number of variance parameters would need to be included in the two-part model.  While such complexities may further complicate inference and increase the computational burden, the authors note that it is important to make these accommodations: ``failing to account for intra-cluster heterogeneity may lead one to substantially overstate the actual precision of the estimates.''

\section{LLT rank based approach}

The rank-based test we propose is based on the work of  Lin et al. (2013) (LL\&T) \cite{lin2013estimating}
 for the analysis of joint time-to-event and quantitative data.  Joint outcomes are often observed in survival analysis settings, and  there are often clear advantages to considering both outcomes together as a composite-endpoint when determining treatment potential  \cite{freemantle2003composite, van2003use, thompson2011survival}.   
 The issue of how to best analyze joint-outcome data of this type has received considerable attention in the literature within the last two decades and a wide range of strategies have been recommended
 \cite{kurland2009longitudinal}.  
  
A reasonable and rather simple option is that of non-parametric rank analysis.  If one can use both time-to-event and quantitative data to compare subjects relative to one another in terms of outcome severity, then the rank of ordered subjects can be used as a single summary measure for the outcome.    (It is important to note that the alternative being tested in all rank based tests is a difference in medians between the two populations and not a difference in the means.)

Moy\'e et al. (1992) \cite{moye1992analysis} 
proposed a simple non-parametric rank method based on a $U$-statistic consisting of pairwise comparisons.  Building on this idea, Finkelstein and Schoenfeld (1999) \cite{finkelstein1999combining} proposed a widely used `joint-rank' method for composite outcome data.  The Finkelstein and Schoenfeld (1999) \cite{finkelstein1999combining} 
method is attractive as it provides, in a simple manner, a single summary statistic based on the pairwise comparisons of subjects.  Simply put, if this statistic is significant and positive, then one can conclude that subjects receiving treatment obtain significantly better (higher ranking) outcomes than their peers.   The proposal of Lin, Li and Tan (2013) \cite{lin2013estimating}
(LL\&T)  is a similar `pair-wise comparison' method that allows for the inclusion of additional covariates and considers all data collected on each subject.

In the following, consider $p$ covariates of interest (with $p \ge 2$) and $\boldsymbol{\beta}$, a $p$-length vector of coefficients.  Note that no intercept term is included as ranks are invariant to location shift.  As in Section 2, for the $i$th observation at the $j$th time-point, let $Y_{ij}$ be the observed outcome and $\boldsymbol{X_{ij}}$ be a $1$ by $p$ matrix of covariates (i.e. no intercept), for $i=1,...,n$ and $j=1,...,m_{i}$.  In the joint time-to-event and quantitative data setting, the observed outcomes, $Y_{ij}$, could be either survival times or quantitative measures.

The LL\&T approach involves maximizing the following objective function, based on the maximum rank correlation (MRC) originally proposed by Han (1987) \cite{han1987non} (see more recently \cite{han2017provable}):

\begin{equation}
S(\boldsymbol{\beta}) =  \frac{1}{n(n-1)}\sum_{i=1}^{n}\sum_{j=1}^{m_{i}}\sum_{k=1}^{n}\sum_{l=1}^{m_{j}}  {I}(Y_{ij}>
Y_{kl}) {I}(\boldsymbol{X^{T}_{ij}}\boldsymbol{\beta}  > \boldsymbol{ X^{T}_{kl}}\boldsymbol{\beta}) \quad , \quad 
\label{eq:mrc}
\end{equation}

\noindent where, for identifiability, $||\boldsymbol{\beta}||=1$.  In the joint time-to-event and quantitative data setting, the ranking of different types of outcomes requires additional considerations for defining ${I}(Y_{ij}>Y_{kl})$; and censored survival times may add additional complications \cite{lin2013estimating}.  In our application with semicontinuous outcomes, all outcomes are of the same type (and measured on the same scale) and ranking the outcomes is therefore straightforward.

Maximizing $S(\boldsymbol{\beta})$  is equivalent to maximizing Kendall's tau correlation coefficient between vectors $\boldsymbol{y}$ and $\boldsymbol{X^{T}\beta}$; see  \cite{han1987non}.  Intuitively, this optimization is looking to obtain an estimate of $\boldsymbol{\beta}$ such that, whenever we have $Y_{ij} \ge Y_{kl}$, it is likely that we also have $\boldsymbol{X^{T}_{ij}}\boldsymbol{\beta}  \ge \boldsymbol{ X^{T}_{kl}}\boldsymbol{\beta}$.  In other words, we wish to have an estimated $\boldsymbol{\beta}$ such that the ranking of the \emph{observed} outcomes is in agreement with the ranking of the \emph{fitted} outcomes.  Ties are irrelevant and are therefore discarded, see Sherman (1993) \cite{sherman1993limiting}.

Unfortunately, due to the discontinuity of the second indicator function which ranks the fitted values, ${I}(\boldsymbol{X^{T}_{ij}}\boldsymbol{\beta}  > \boldsymbol{ X^{T}_{kl}}\boldsymbol{\beta})$, the $S(\boldsymbol{\beta})$  is a discontinuous step function with abrupt changes.  This makes optimization difficult, particularly in high dimensions (i.e. when $p$ is large).  The most common approach to overcome this sort of challenge is to approximate the discontinuous $S(\boldsymbol{\beta})$ function with a smooth substitute function (e.g. using a sigmoid function \cite{ma2005regularized}, or using a standard Gaussian cdf  \cite{lin2013smoothed}).   Other approaches involve clever, yet computationally costly, grid-search type optimization techniques, see Wang (2007) \cite{wang2007note}.  LL\&T make use of smoothing with the standard Gaussian cdf, $\Phi()$, such that the estimated $\boldsymbol{\beta}$  is defined as:
\begin{equation}
\boldsymbol{\hat{\beta}} = \textrm{argmax}_{\boldsymbol{\beta}} \frac{1}{n(n-1)}\sum_{i=1}^{n}\sum_{j=1}^{m_{i}}\sum_{k=1}^{n}\sum_{l=1}^{m_{j}}  {I}(Y_{ij}>Y_{kl}) \Phi((\boldsymbol{X_{ij}^{T}}\boldsymbol{\beta}  - \boldsymbol{ X_{kl}^{T}}\boldsymbol{\beta})/h) ,
\label{eq:mrc2}
\end{equation}


\noindent where $h$, the bandwidth, is a very small constant that converges to zero as the sample size, $n$, increases.  Based on the work of Lin and Peng (2013) \cite{lin2013smoothed} (see also \cite{han1988large, sherman1993limiting}),  LL\&T show that, given a sufficiently small value for $h$ (we require $nh^{4} \rightarrow 0$) and a finite number of observations per subject (we require $\textrm{max}_{i}(m_{i}) < \infty$), and five other regularity conditions, the estimate $\boldsymbol{\hat{\beta}}$  is $\sqrt{n}$-consistent and is asymptotically normal \cite{lin2013estimating}.  As such, LL\&T declare that the selection of $h$ is ``not crucial for the asymptotic performance of the estimate'' \cite{lin2013estimating, lin2011selection}. (See also, Lin and Peng (2013) \cite{lin2013smoothed}: ``our method is not sensitive to the bandwidth parameter $h$''.)   LL\&T  \cite{lin2013estimating} suggest taking $h$ equal to $\tilde{\sigma}/n^{1/3}$ , where $\tilde{\sigma}$ is the approximated sample standard deviation of $\boldsymbol{X^{T}\hat{\beta} }$ (and where $\boldsymbol{\hat{\beta}}$ can be estimated by iterating over equation (\ref{eq:mrc2}) with some initial value chosen for $h$).

In order to determine the statistical significance of $\boldsymbol{\hat{\beta}}$, its sampling distribution is approximated via non-parametric resampling.  LL\&T make use of the simple and clever resampling technique of Jin et al. (2001) \cite{jin2001simple}, whereby a statistic is perturbed with a large number ($B$) of independent draws from a random variable with mean 1 and variance 1. As Zhou et al. (2005) note, ``it is not clear what is the preferred distribution for the perturbation''  \cite{zhou2005empirical}.  As such, we will use exponential perturbations, a relatively popular choice (e.g. \cite{peng2008survival, jin2003rank, lin2013estimating, fan2017concordance}).  Note that, in our implementation, perturbations to the $\boldsymbol{\hat{\beta}}$ statistic are made at the subject-level so as to account for the correlation between repeated measurements.  (As an alternative to the perturbation resampling technique, a bootstrap resampling scheme could be implemented \cite{subbotin2007asymptotic, field2007bootstrapping}.  In fact, the resampling technique of Jin et al. (2001) \cite{jin2001simple} is known to be a special case of Rubin (1981)'s `Bayesian bootstrap', see \cite{parzen2007perturbing, rubin1981bayesian}.  In a bootstrap-based implementation, in order to  account for the correlation between repeated measurements, data must be resampled at the subject level.)
 
What follows is a step-by-step summary of the LL\&T algorithm implemented for longitudinal semicontinuous outcome data in two parts. 
 
\vspace{0.75cm}

\noindent \underline{The LL\&T algorithm for longitudinal semicontinuous outcome data.}{}%
\\
\textbf{Part 1- Establishing the bandwidth and point estimate.} Define $\tilde{\sigma}^{[1]}=1$. \\
Then, for $q$ in 1,...,$Q$ (or until $\tilde{\sigma}^{[q]} \approx \tilde{\sigma}^{[q-1]}$):

\hangindent=1.5cm 1. With $\Phi()$ as the Normal cdf, and with the restriction that $||\boldsymbol{\beta}||=1$, use numerical methods to maximize:  
\begin{equation*}
\boldsymbol{\hat{\beta}^{[q]}} = \textrm{argmax}_{\boldsymbol{\beta}} \frac{1}{n(n-1)}\sum_{i=1}^{n}\sum_{j=1}^{m_{i}}\sum_{k=1}^{n}\sum_{l=1}^{m_{j}} {I}(Y_{ij}>Y_{kl}) \Phi((\boldsymbol{\beta^{T}} X_{ij} - \boldsymbol{\beta^{T}} X_{kl})/h^{[q]})
\end{equation*}

\hangindent=1.5cm 2. Define $\tilde{\sigma}^{[q+1]}$ as the standard deviation of $\boldsymbol{X}^{T}\boldsymbol{\hat{\beta}^{[q]}}$.

\hangindent=1.5cm 3. Define  $h^{[q+1]}=\tilde{\sigma}^{[q+1]}/n^{1/3}$.

\noindent \textbf{Part 2- Calculating $p$-values.} Take the point estimate, $\boldsymbol{\hat{\beta}}$, and the bandwidth, $h$, from the final iteration of Part 1.\\
Then, for $b$ in 1,...,$B$:

\hangindent=1.5cm  1. For $i$ in 1,...,$n$, sample $\zeta_{i}^{[b]}$ from the exponential distribution with mean and variance equal 1. 

\hangindent=1.5cm 2. {With $\Phi()$ as the Normal cdf, and with the restriction that $||\boldsymbol{\beta}||=1$, use numerical methods to maximize:  
\begin{equation*}
\boldsymbol{\beta^{[b]}} = \textrm{argmax}_{\boldsymbol{\beta}} \frac{1}{n(n-1)}\sum_{i=1}^{n}\sum_{j=1}^{m_{i}}\sum_{k=1}^{n}\sum_{l=1}^{m_{j}} \zeta_i^{[b]}\zeta_j^{[b]} {I}(Y_{ij}>Y_{kl}) \Phi((\boldsymbol{\beta^{T}} X_{ij} - \boldsymbol{\beta^{T}} X_{kl})/h).
\end{equation*}}

\vspace{-0.8cm}

\noindent {Define one-sided $p$-values, for $j$ in 1, $...,p$, as follows:   $p\textrm{-value}_{j} =\frac{1+\sum_{b=1}^{B}I(\beta_{j}^{[b]}>0)}{(1+B)}.$}\\
 {Define two-sided $p$-values,  for $j$ in 1, $...,p$, as follows: \\
$p\textrm{-value}_{j} = 2 \cdot min\left(\frac{1+\sum_{b=1}^{B} I(\beta_{j}^{[b]} > 0)}{1+B}, \frac{1+\sum_{b=1}^{B} I(\beta_{j}^{[b]} \le 0)}{1+B} \right).$

}

\vspace{0.75cm}


Four comments are necessary to clarify the steps outlined above.{ First, the constrained maximization required can be achieved by first transforming the $\boldsymbol{\beta}$ vector of length $p$ into a vector $\boldsymbol{\theta}$ of length $p-1$ of corresponding polar coordinates.  For example, with $p=2$,  simply define $\boldsymbol{\beta} = (sin(\boldsymbol{\theta}), cos(\boldsymbol{\theta}))$.  Transformations for higher dimensional polar coordinates can be obtained similarly \cite{nolan2015sphericalcubature}. Maximization with respect to the vector $\boldsymbol{\theta}$ can then be achieved using standard optimization methods (e.g. Newton-type algorithms).  Alternatively, one could maximize with respect to the untransformed $\boldsymbol{\beta}$ vector using available techniques for optimization subject to nonlinear constraints (e.g. \cite{chen2017nlcoptim}).}  

Second, in the event of ties (i.e. equal ranks, $Y_{ij}=Y_{kl}$), the indicator function equals zero, (i.e. ${I}(Y_{ij}>Y_{kl})=0$), see  \cite{sherman1993limiting}.  Third, note that the coefficient estimates are scaled ($||\boldsymbol{\beta}||=1$) so that the norm of the estimates is always equal to 1.  While this may not be ideal for interpretation, it is necessary to maintain identifiability.  We recommend this test only for data with $p\ge2$.  If $p=1$ (i.e. only one covariate is considered), the only two possibilities for $\boldsymbol{\beta}$ are values of $-1$ and $1$ and the sampling distribution of the $p$-value will be poorly behaved.  Finally, we note that, for effective Newton-type optimization, one should consider multiple different initial values so as to avoid local maxima.   In our simulation study and application, we select six random initial values as starting points.  For each of these six points, we run three Newton-Raphson steps and continue forward with Newton-Raphson optimization from the point for which the objective function is greatest.

\section{Simulation Study}

A small simulation study was conducted comparing the power and type I error of four methods.  The simulations were coded in SAS in order to make use of existing procedures and macros.  We compared the following four approaches for hypothesis testing:

\begin{enumerate}
\item{a mixed logistic model (using PROC GLIMMIX)};
\item{the two-part model (with $g(Y)=log(Y)$) of O\&S \cite{olsen2001two} (using PROC NLMIXED as in \cite{liu2012analyzing, tooze2002analysis}) ;}
\item{the mixed Tobit model with a log-transformed outcome (i.e. $z=log(Y),\textrm{ if }Y>0; z=0,\textrm{ otherwise})$ (using PROC NLMIXED) \cite{twisk2009longitudinal}; and}
\item{the proposed LL\&T non-parametric rank based test (using PROC IML), with $B=101$ and $Q=5$.}
\end{enumerate}

 Data was simulated from the random intercept model (see equations 1-3) based in part on the simulation work of O\&S \cite{olsen2001two}.  The total number of subjects was equal to $n$ and the number of measurements per subject was varied with $m_{i}$ = 5 + $p_{i2}$, where $p_{i2}$ $\sim$ Poisson($\lambda$=2), for $i$ in 1 to $n$.    For each subject, $i=1,...,n$, we generated a non-time-varying covariate $(X_{1})_{i}$ from $Bernoulli(0.5)$.  This $X_{1}$ was the ``variable of interest''.  We also defined a time-varying covariate, $X_{2}$,  as $(X_{2})_{ij}=j$, for $j=1,...,m_{i}$.  We then generated the $V_{i}$s from a $N( X^{*}_{i}\gamma+ Z^{*}_{i}d_{i}, \sigma^{2}I)$  distribution and $U_{i}$s  from a $Bernoulli(expit(X_{i}\beta +Z_{i}c_{i}))$ distribution.  Finally, we implemented two different scenarios:

\begin{itemize}
\item{Scenario 1 :  We define $Y_{ij}=0$  if ($U_{ij}=0$ or $V_{ij}\le0$), and $Y_{ij}=exp(V_{ij})$ otherwise.}

\item{Scenario 2 :  We define $Y_{ij}=0$  if ($U_{ij}=0$ or $V_{ij}\le0$), and $Y_{ij}=exp(\sqrt{V_{ij}})$ otherwise.}
\end{itemize}

As such, the parametric assumptions required of the two-part model and Tobit model (that the function $g()$ is such that $V_{ij}$ is normally distributed) will be satisfied for Scenario 1 but not quite satisfied for Scenario 2.  {In practice, upon observing the data, one could conceivably use a monotone function to transform the $V_{ij}$ variable as needed.  However, for the purposes of this simulation study, we will not allow this strategy as we are interested in investigating the effects of violations to the parametric assumptions (or equivalently, the misspecification of the $g()$ function).}

Parameter values for $\beta_{0}$, $\gamma_{0}$, $\beta_{2}$, $\gamma_{2}$, $\sigma^{2}$, $\psi_{cc}$, $\psi_{cd}$ and $\psi_{dd}$ were fixed at: $\beta_{0} =0.25$, $\gamma_{0} =2.5$,  $\beta_{2}=0.15$, $\gamma_{2}=0.15$, $\sigma^{2} =0.25$, $\psi_{cc} =0.0625$, $\psi_{cd} =0.0625$, $\psi_{dd} =0.065$.  Sample size took one of three values, with $n$ = 50, 100 or 150 subjects.  Under these settings, approximately 30\% of observations were equal to 0.

 To study type I error,  $\beta_{1}$ and $\gamma_{1}$ were set equal to zero.  To study power, $\beta_{1}$  took values 0.1 and 0.25 and $\gamma_{1}$ took values 0.1 and 0.25.  For each combination of  settings, 500 simulated datasets were generated.   The empirical estimate of statistical power is the proportion of simulated datasets in which convergence is achieved and the null hypothesis is rejected (two-sided alternative) at $\alpha$-level 0.05.  Parameter values and sample sizes for the simulations were chosen after a small pilot run of the simulation study suggested they would result in a sufficiently wide range of observed statistical power.  Details on the four methods and their implementation are available in the Appendix.  Also in the Appendix is the SAS code used to generate the data.  SAS Code to implement the four methods is available upon request from the author.

\begin{table}
\tbl{Scenario 1-- Simulations study results show power to detect a significant effect at the $\alpha$=0.05 level.  Data conforms to the parametric assumptions of the two-part model. The empirical estimate of statistical power is the proportion of simulated datasets in which convergence is achieved and the null hypothesis is rejected (two-sided alternative) at $\alpha$-level 0.05.  Results are based on 500 simulations runs. Numbers displayed in \textbf{bold} indicate {best-}performing method (within $\pm$ 1\%) for a given scenario.}{
\begin{tabular}{lcccccccc} \toprule
  $n$ &  $\beta_{1}$  & $\gamma_{1}$  &  rank  & Tobit & two-part  & logistic  & two-part conv.   & logistic  conv.\\ 
 & & & test & model & model &model  &   failures (\%) &   failures (\%)\\
  \hline
 50 & 0 & 0 & 0.07 & 0.06 & 0.02 & 0.09 & 62.40 & 1.20 \\   
  100 & 0 & 0 & 0.04 & 0.07 & 0.02 & 0.05 & 60.40 & 1.80 \\ 
 150 & 0 & 0 & 0.04 & 0.05 & 0.02 & 0.05 & 59.60 & 0.60 \\  
  \hline
50 & 0.25 & 0.10 & \textbf{0.19} & \textbf{0.20} & {0.04} & {0.12} & 65.60 & 0.60 \\  
 100 & 0.25 & 0.10  & \textbf{0.36} & 0.30 & {0.08} & 0.28 & 58.00 & 0.40 \\   
150 & 0.25 & 0.10 & \textbf{0.54} & 0.48 & {0.09} & 0.39 & 65.20 & 0.60 \\ 
 50 & 0.25 & 0.25 & 0.53 & \textbf{0.71} & 0.15 & {0.18} & 65.40 & 1.20 \\ 
 100 & 0.25 & 0.25 & 0.83 & \textbf{0.95} & 0.18 & {0.28} & 69.00 & 0.40 \\ 
 150 & 0.25 & 0.25 & 0.94 & \textbf{0.98} & 0.23 & 0.40 & 71.60 & 0.40 \\  
 50 & 0.10 & 0.25 & 0.40 & \textbf{0.73} & 0.20 & {0.05} & 66.60 & 1.20 \\  
 100 & 0.10 & 0.25 & 0.66 & \textbf{0.95} & 0.24 & {0.10} & 70.20 & 1.60 \\
 150 & 0.10 & 0.25 & 0.86 & \textbf{0.99} & 0.28 & {0.12} & 70.00 & 0.20 \\ 
\hline
\end{tabular}}
\begin{tabnote}
\end{tabnote}
\end{table}

\begin{table}
\tbl{Scenario 2-- Simulations study results show power to detect a significant effect at the $\alpha$=0.05 level. Data does not conform to the parametric assumptions of the two-part model.  Results are based on 500 simulation runs.  The empirical estimate of statistical power is the proportion of simulated datasets in which convergence is achieved and the null hypothesis is rejected (two-sided alternative) at $\alpha$-level 0.05.  Numbers displayed in \textbf{bold} indicate {best-}performing method (within $\pm$ 1\%) for a given scenario.  }{
\begin{tabular}{lcccccccc} \toprule
  $n$ & $\beta_{1}$ & $\gamma_{1}$ & rank  & Tobit & two-part  & logistic  & two-part conv.   & logistic  conv.\\ 
 & & & test & model & model &model  &   failures (\%) &   failures (\%)\\
\hline
  50   & 0 & 0 & 0.05 & 0.05 & 0.01 & 0.05 & 89.80 & 0.20 \\ 
 100 & 0 & 0 & 0.03 & 0.06 & 0.00 & 0.05 & 94.20 & 0.60 \\ 
 150 & 0 & 0 & 0.04 & 0.06 & 0.00 & 0.05 & 91.80 & 2.00 \\ 
  \hline
 50   & 0.25 & 0.10 & \textbf{0.21} & {0.16} & {0.03} & {0.16} & 87.40 & 0.20 \\  
 100 & 0.25 & 0.10 & \textbf{0.43} & 0.33 & {0.02} & {0.27} & 96.00 & 0.20 \\ 
 150 & 0.25 & 0.10 & \textbf{0.50} & {0.38} & {0.02} & {0.39} & 95.60 & 1.00 \\
 50   & 0.25 & 0.25 & 0.52 & \textbf{0.71} & {0.08} & {0.14} & 87.80 & 1.00 \\  
 100 & 0.25 & 0.25 & 0.84 & \textbf{0.95} & 0.03 & {0.31} & 95.60 & 0.80 \\  
  150 & 0.25 & 0.25 & 0.95 & \textbf{0.99} & 0.06 & {0.39} & 92.80 & 0.20 \\
 50   & 0.10 & 0.25 & 0.42 & \textbf{0.74} & 0.09 & {0.06} & 85.20 & 1.00 \\ 
  100 & 0.10 & 0.25& 0.66 & \textbf{0.94} & 0.06 & {0.07} & 92.80 & 1.60 \\ 
  150 & 0.10 & 0.25 & 0.82 & \textbf{1.00} & 0.06 & {0.09} & 92.80 & 1.20 \\ 
\hline
\end{tabular}}
\end{table}

\subsection{Simulation study results}

Results of the simulation study are presented in Table 2 and Table 3.  Before reviewing the results with regards to statistical power, consider simulations under the null, when the true $\beta_{1} = 0$ and $\gamma_{1} = 0$.   Several findings merit comment.

\begin{itemize}
\item {For both scenarios, the rank, Tobit, and logistic approaches show relatively desirable type I error of approximately 0.05 even when sample size is small ($n = 50$). It is difficult to evaluate the type 1 error for the two-part model, given that convergence is not achieved in a majority of simulations.}
\item{For scenario 1, the two-part model fails to converge in upwards of 60\% of simulations.  This is in line with the reported findings of O\&S \cite{olsen2001two}.  For scenario 2, the two-part model fails to converge in approximately 90\% of simulations.  This suggests that the two-part model cannot accommodate even modest violations to the stated parametric assumptions.}
\end{itemize}

With regards to statistical power, we obtain a wide range of power across methods which varies considerably with the set values of $\beta_{1}$ and $\gamma_{1}$.  Consider the following. For \textbf{Scenario 1} (see Table 2):
\begin{itemize}
\item {when $\beta_{1}=0.25$ and $\gamma_{1}=0.10$, the rank test appears to have the highest power relative to the other methods.}
\item {when $\beta_{1}=0.25$ and $\gamma_{1}=0.25$, the power of the logistic model remains relatively low, while the other methods all show higher power;  the Tobit model shows higher power than the rank test which has higher power than the two-part model.}
\item {when $\beta_{1}=0.10$ and $\gamma_{1}=0.25$, the power of the logistic model is negligible; the Tobit model shows higher power than the rank test and the two-part model.}
\item {for all settings, the rank test and the Tobit model show higher power than the logistic model and the two-part model. }
\end{itemize}
For \textbf{Scenario 2} (see Table 3):
\begin{itemize}
\item {the two-part model fails to converge for more than 85\% of simulated datasets.     }
\item {when $\beta_{1}=0.25$ and $\gamma_{1}=0.10$, the rank test appears to have the highest power relative to the other methods.}
\item {when $\beta_{1}=0.25$ and $\gamma_{1}=0.10$, the Tobit model has about the same  power as the logistic model; when $\beta_{1}=0.10$ and $\gamma_{1}=0.25$, the Tobit model shows the highest power; and also when $\beta_{1}=0.25$ and $\gamma_{1}=0.25$, the Tobit model shows the highest power. }
\item {for all settings, the rank test shows higher power than the logistic model.  The same can not be said of the Tobit model.}
\end{itemize}

Given the relatively successful performance of the random-effects Tobit model, further discussion of this method is appropriate.  The likelihood of a Tobit model explicitly incorporates both the probability that an observation is below or equal to zero and the probability distribution of an observation conditional on the fact that it is above zero \cite{su2017analysis}.  This could explain why the Tobit model appears rather efficient for testing semicontinuous data.  In our simulations, when the assumption of normality was invalid (i.e. in Scenario 2), we still observed relatively high power with the Tobit model under most settings.  However, based on other results in the literature, we remain skeptical as to whether the Tobit model is robust in all such circumstances and under other departures from normality; see \cite{barros2018generalized, holden2011testing}.  After a thorough investigation, Arabmazar and Schmidt (1982) \cite{arabmazar1982investigation} note of the Tobit model that : ``The bias from non-normality can be substantial'' and that the ``bias due to non-normality depends on the degree of censoring.'' For our simulated semicontinuous data, the proportion of observations equal to zero was fairly consistent at approximately 30\%.

\section{Motivating Example: Is it even rainier in \emph{North} Vancouver?}

Canadians love to complain about the weather, and those in Vancouver on the {We(s)t}-coast, are no exception due to the never-ending rainfall.  From October to March, it is not uncommon to have three soaking weeks worth of non-stop cold and rainy weather.  The daily downpours have precipitated, among umbrella-wielding Vancouverites, a \emph{not-so}-heated debate as to whether neighbouring North Vancouver is even rainier.  \emph{Across the water is the weather even wetter?} Environment Canada's historical weather data  is publicly available (www.climate.weather.gc.ca/) and can help us answer this pertinent pluviometric query.

Using four recent years (2013-2016) of daily data from two weather stations (`Vancouver Harbour CS' and `N. Vancouver Wharves' at a distance of only 2.23km from one another) we can illustrate how each statistical method compares in terms of efficiency:  how many weeks of data are required to correctly conclude that the city North Vancouver is indeed rainier (i.e. reject the null hypothesis)?  Table \ref{table:raindatatable} provides a summary splash of the data.  (One outlier daily observation (N.Van., 2015.11.30) that was obviously a typo was dropped.)

We randomly sampled 50 (75, and 100) week-city pairs worth of observations and tested for a difference in rainfall between the two cities using the four statistical methods as in the simulation study.  The models used in the simulation study were fit nearly identically to the rainfall data (one difference was setting $B=201$ for added precision).  Each week served as an observational unit with days within a week correlated.  The $X_{1}$ variable specified the city ($X_{1}$ = 0 : `Vancouver'; $X_{1}$ = 1 : `North Vancouver'). The $X_2$ variable was set as an indicator of the season such that $X_2$ = 0 represented `April to September' and $X_2$ = 1 represented `October to March.' 

Due to missing data, not all weeks included seven days.  No effort was made to force balance (i.e. in a given dataset, the number of observed days from Vancouver and North Vancouver was not set to be equal, nor was the timing of these observed days).  We repeated the exercise 100 times, recording all two-sided $p$-values. We then evaluated the performance of each model by measuring the proportion of sampled datasets for which the method achieves convergence and correctly rejects the null hypothesis.  In this way, we are able to evaluate, to a certain degree, the efficiency of the methods with this real-world data.  However, do note that this evaluation of efficiency is imperfect: since we are randomly sampling observations from a finite dataset (four years of data), each of the 100 evaluations cannot be considered independent.

Table \ref{table:raintable} shows the results in terms of the number of times in which each method converged and rejected the null hypothesis.  While by no means a deluge, the Tobit model rejected the null the most: in 28 (39, and 55) out of 100 instances.  The rank test did not perform quite as well, rejecting the null 6 (12, and 44) times out of 100.  In contrast, the logistic model shows negligible efficiency, while, for the majority of samples, convergence failures {overcast} the two-part model.  Clearly, neither the logistic model nor the two-part model can be recommended for distinguishing the drizzly days.  The number of times in which the two-part model both converges and successfully rejects the null, is like that of sunny days in the winter months, only 7 (5, and 10) out of 100. It is worth noting that both covariates in this data analysis are binary. This may be one reason for the relatively limited performance of the rank-based test.  If $X_1$ and $X_2$ are exactly the same for two sample units, then all corresponding pairwise comparisons will contribute nothing to the estimation of $\boldsymbol{\beta}$ (see equation \ref{eq:mrc2}).  The rank based test may be best suited for data in which at least one covariate is continuous.

\begin{table}
\tbl{Rainfall Data Summary-- Summary of daily rainfall data (2013-2016) obtained from Environment Canada on two weather stations, `Vancouver Harbour CS' and `N. Vancouver Wharves'; `rainy season' is October to March.}{
\begin{tabular}{lcccccc} \toprule
 & Number of  $\quad$ & Total mm $\quad$  & ...during rainy $\quad$ & Daily mean  $\quad$  & Daily Max. $\quad$ & NA's \\ 
 &  non-rainy days $\quad$ & recorded $\quad$ & season (mm) $\quad$ &   (mm) $\quad$ &  (mm) $\quad$ &  (days)\\ 
  \hline
Vancouver & 748 & 4296.00 & 3242.60  & 3.20  &  46.60  & 114 \\ 
North Vancouver & 657 & 5851.50  & 4367.30  & 4.58 & 64.00  & 183 \\ 
   \hline
\end{tabular}}
\label{table:raindatatable}
\end{table}

\begin{table}
\tbl{Rainfall Data-- Results show number of analyses (out of 100) in which convergence is achieved and a significant effect at the $\alpha$=0.05 level is detected (i.e. in which the null hypothesis is rejected).}{
\begin{tabular}{lcccccc} \toprule
  $n$  & rank  & Tobit & two-part  & logistic  & two-part model   & logistic model \\ 
& test & model & model &model  &  conv. failures  &   conv. failures \\
  \hline
  50 & 6 & 28 & 7 & 5 & 68 & 0 \\ 
    75 & 12 & 39 & 5 & 9 & 61 & 2 \\ 
   100 & 44 & 55 & 10 & 11 & 51 & 0 \\ 
   \hline
\end{tabular}}
\label{table:raintable}
\end{table}

\section{Conclusion}

How to best address the challenges of longitudinal semicontinuous data continues to be an important and difficult statistical question.  Neelon et al. (2016) \cite{neelon2016modeling} provide a summary of the current state of affairs and related issues concerning zero-inflated count data. 

In this work, we proposed using an existing testing method for a different application than the one for which it was originally intended.  The LL\&T rank-test has been proposed and recommended only for applications of joint longitudinal survival data.  We believe it has far greater potential.  To the best of our knowledge, using a rank-based method for testing longitudinal semicontinuous data has not yet been considered in the literature (one exception might be \cite{tran2015assessing} who very briefly consider the standard Wilcoxon rank sum test).  This is no doubt due to the fact that standard rank tests cannot incorporate multiple (both continuous and catagorical) covariates; nor can they adjust for the correlation amongst longitudinal (or clustered) observations.  Recently, some progress has been made in these regards; see e.g. \cite{datta2008signed, jiang2017wilcoxon}. 

 We believe the LL\&T rank-test has much potential for  longitudinal semicontinuous data (as well as other data with unorthodox distributions) as it allows for multiple covariates and correlated observations in  a straightforward way.  Furthermore, as we established in the simulation studies, the LL\&T test shows relatively high efficiency.  That being said, it remains to be determined how well this rank based test will work with a larger number of covariates, $p$, and a larger number of observations, $n$.  One thing is certain, optimization of the $S(\beta)$ objective function will be much more computationally costly with larger values of $p$ and $n$; see the recent work of Fan et al. (2017) \cite{fan2017rank}.  While the computational cost of the rank-test can be substantial, it may still remain feasible in many applications due to the fact that the algorithm is easily parallelized.

Our simulation study suggested that the advantages, with regards to efficiency, of using the rank-test over alternatives such as the random effects Tobit model, will be dependent on whether the effect is primarily expressed through the binary or continuous components of the data (i.e. will be dependent on the relative magnitude of $\beta_{1}$ and $\gamma_{1}$).  Among the four methods tested, the LL\&T rank-test is the most powerful in situations when a substantial amount of the treatment effect is expressed through the binary component of the data (i.e. when the magnitude of $\beta_{1}$ is large relative to the magnitude of $\gamma_{1}$).  

Our simulation study also confirmed that computational issues can greatly restrict the use of the two-part model in many cases.  In fact, we observed that when distributional assumptions of the two-part model do not hold, model convergence is almost always unattainable.  Should computational issues be resolved, the two-part model would be, without a doubt, a most desirable approach.  Current research on this is promising (e.g. \cite{zhang2016composite} and the Bayesian methods of \cite{dreassi2017bayesian}).  For those seeking a simpler approach, the random-effects Tobit model and the LL\&T rank-test are good options to consider.  

Previous concerns about the suitability of the random-effects Tobit model appear largely unwarranted based on our results (at least with regards to testing).  Indeed, the performance of the Tobit model in our simulation study does seem at odds with recommendations in the literature \cite{schafer1999modeling, olsen2001two}.  Further research on this  is warranted.  If the distribution of the continuous outcome appears particularly skewed (even after transformations) such that the assumption of normality may not hold, new research into skew-normal Tobit models may prove useful \cite{su2017analysis}.  With respect to the LL\&T rank-test, testing is efficient while interpretation is less straightforward due to the fact that coefficient estimates are scaled. Our simulation study also confirmed that truncating semi-continuous data to binary can be very detrimental in terms of power.  In all cases considered, the LL\&T rank-test was preferable to the simple mixed logistic model approach.

In many circumstances when testing is the primary objective, the choices required to fit parametric models (e.g. `what random-effects to include?', `what function $g()$ is most appropriate?') can be difficult to justify and the many model assumptions can prove burdensome.  In such a setting, a main advantage for the LL\&T rank-test is the lack of parametric assumptions and required modelling choices.  

{The data considered in this paper were chosen to be the simplest possible in order to best illustrate the main challenges involved with longitudinal semicontinuous data.  As such, not all aspects of the data were considered in the analysis.  For example, we treated each city-week as an independent unit, ignoring any residual correlation between adjacent weeks for the same city.  This type of issue and other difficulties may often arise in practice.}  For example, what to do about informative cluster sizes (i.e. when the number of observations is related to the outcome) or missing data?  How to best accommodate serial correlation (e.g. time-series data)?  There are many potential areas for future methodological research.  First and foremost, further investigation  of the rank based test with regards to its sensitivity to the selection of the bandwidth is required.  While the choice of bandwidth is ``not crucial for the asymptotic performance of the estimate'' \cite{lin2013estimating}, it may have the potential to impact the statistical power of the test.  In related work, Horowitz (2002) \cite{horowitz2002bootstrap} investigates, for tests based on smoothed maximum score estimators, how bootstrap critical values are sensitive to the choice of the bandwidth.  Less is known about the impact of bandwidth selection when using the perturbation resampling method of of Jin et al. (2001) \cite{jin2001simple}.

Finally, with the daily Vancouver rainfall data, our goal was not to properly model the meteorological process (as others have done, e.g. \cite{fleming2007climatic, george2016daily}).  Rather we would hope that such a simple example, in which established methods show a drought of efficiency and an inundation of computational issues, might {precipitate} the use alternative methods.

\section*{Acknowledgements}
Thank you to Dr. Kahlil Baker for the friendly and thought-provoking computational assistance.  Thank you to Professors Lang Wu, Nancy Heckman, and Paul Gustafson for the thoughtful discussions and suggestions.
------

\section*{Notes}

Note that this work has been previously posted as a pre-printed, see \cite{campbell2017even}.  In addition, the rain data and the sas code for the simulation study has been posted to the publicly available OSF repository (DOI 10.17605/OSF.IO/PZGSA).  Finally, to promote the use of the LL\&T rank-test we encourage those interested to make use of the \texttt{R} function \texttt{LLTranktest} available in the Appendix C and within a package on the Github repository ``harlanhappydog/LLTranktest''.  The following four lines of  \texttt{R} code will download and install the ``LLTranktest'' package, and use the LL\&T rank-test to test for a treatment effect in the analysis of the longitudinal semicontinuous ``toenail data'' \cite{backer199612}, also analyzed in \cite{mian2012two} and \cite{mahabadi2014bayesian}:

\begin{verbatim}
library(devtools)
install_github("harlanhappydog/LLTranktest")
library(LLTranktest)
LLTranktest(UNL_mm~treat_group+month, id=toenail$ID, data=toenail, B=1000)
\end{verbatim}

\bibliographystyle{tfs}
\bibliography{semicont-refs}

\appendix

\section{Simulation study details}

\paragraph{\textbf{--The Random effects Tobit model}}  First, a logarithmic transformation is made, such that $z_{ij}=log(Y_{ij}) \textrm{ if } Y_{ij}>0;  z_{ij}=0$ otherwise, for $i=1,...,n$ and $j=1,...,m_{i}$. As such, the normality assumption is satisfied for Scenario 1 but not so for Scenario 2.  Then, using the SAS NLMIXED procedure (with 5 adaptive quadrature points) the likelihood function is maximized:
\begin{align*}
 L &= ({{\sigma^{2} \sqrt {2\pi } }})^{-1/2}exp({{{ - \left( {z_{ij} - \mu_{ij} } \right)^2 } \mathord{\left/ {\vphantom {{ - \left( {x - \mu } \right)^2 } {2\sigma ^2 }}} \right. \kern-\nulldelimiterspace} {2\sigma ^2 }}}) , \quad \textrm{if} \quad z_{ij}>0,  \\
&=\int_{-\infty}^{0} ({{\sigma^{2} \sqrt {2\pi } }})^{-1/2}exp({{{ - \left( {v - \mu_{ij} } \right)^2 } \mathord{\left/ {\vphantom {{ - \left( {x - \mu } \right)^2 } {2\sigma ^2 }}} \right. \kern-\nulldelimiterspace} {2\sigma ^2 }}}) dv  , \quad \textrm{if} \quad z_{ij}=0;\\
\textrm{where:    } &\quad  \mu_{ij} = \beta_{0} + \beta_{1}X_{1ij} + \beta_{2}X_{2ij} + u_{i}
 \quad \textrm{    and     } \quad u_{i} \sim N(0, \sigma_{s}^{2}).
\end{align*}

 Initial values for the random effects Tobit model (PROC NLMIXED) were obtained by first fitting a standard Tobit model (using PROC LIFEREG), and by setting $ \sigma_{s} = \sigma =0.5$.

\paragraph{\textbf{--The Binary logistic model}}The simplest way to approach longitudinal semicontinuous data, is to reduce the data to longitudinal binary data and fit a mixed effects logistic model.  This approach is equivalent to only considering $U$ and ignoring $V$.  The downside to this method is that ignoring $V$ will most certainly result in a reduction of power. Initial values for the logistic mixed effects model (PROC GLIMMIX) were obtained by first fitting a fixed-effects generalized linear model (GLM).

\paragraph{\textbf{--The two-part model}}

SAS macros based on the work of Tooze et al. (2002) and Liu et al. (2012), were coded for all simulations.  Adaptive Gaussian quadrature within the SAS NLMIXED procedure was used (with 5 adaptive quadrature points as in Liu et al. (2010) 
\citep{liu2010flexible}
who note that: `[i]ncreasing the number of quadrature points to 10 substantially increased computation time with negligible changes to the results').   Initial values for the $\gamma_{0}$, $\gamma_{1}$, $\beta_{0}$, and $\beta_{1}$ parameters were obtained from fitting Generalized Estimating Equation (GEE) models (PROC GENMOD) and initial values for $\sigma$, $\psi_{cc}$, $\psi_{cd}$ and $\psi_{dd}$ were all set to equal 0.5.

\paragraph{\textbf{--The non-parametric rank-based test}}  A two-sided $p$-value was obtained as described in steps in Section 3 with $B$ was set at 101 and $Q$ was equal to 5.  This $B$ was much smaller than ideal, due to restricted computational capacity.  The restricted ($||\beta||=1$) maximization was made simpler by use of the trigonometric identity $sin(x)^{2}+cos(x)^{2} =1$.  We set $\beta_{1}=sin(\theta)$ and $\beta_{2}=cos(\theta)$ and then maximized by Newton-Raphson iteration with respect to $\theta$.  In order to avoid simply selecting local optima, we selected six random initial values as starting points.  For each of these six points, we ran three Newton-Raphson steps and continued forward with Newton-Raphson optimization from the point for which the objective function was greatest.

\section{SAS code used to simulate data}
\begin{footnotesize}
\begin{verbatim}
	m = &m;	alpha0=0.25;	alpha1=&alpha1;	beta0=2.5;	
	beta1=&beta1;	sigma = 0.5;	psiaa =0.25;	psibb=0.05;
		do j=1 to m;
			ID = j;
			ni =  5 + rand("Poisson", 2);
			ai = rand("Normal", 0, psiaa);  bi = rand("Normal", ai, psibb);
			xtest  =  rand("Bernoulli", 0.5);
			do i = 1 to ni;
				X2 = i;
   				linpred = alpha0 + xtest * alpha1 + 0.15*X2 + ai;
   				prob = exp(linpred)/ (1 + exp(linpred));
   				ytest = uniform(0) lt prob;
   				logz = exp(rand("Normal", beta0 + xtest*beta1 + 0.15*X2+ bi, sigma));
   				Y = ytest *  logz;
				ZZ = log(Y);
				if zz=. then zz=0;
  				X1 =xtest;
   			output;
   			end;
		end;
		
		
	\end{verbatim}	
	\end{footnotesize}

\section{\texttt{R} code to use LLTranktest}
\begin{footnotesize}
\begin{verbatim}

## null example:
## mydat<-data.frame(Y=rnorm(60), X1=rnorm(60), X2=rnorm(60), X3=rnorm(60), ID=sort(c(rep(1:20,3))))
## LLTrankTtest(Y~X1+X2+X3, id=mydat$ID, data=mydat, B=200)


####################################################
LLTranktest<-function(formula, id=NULL, data, B=200, Q=10){

  require(SphericalCubature)
  ##### The setup
  cl <- match.call()

  mf <- match.call(expand.dots = FALSE)
  m <- match(c("formula", "data"), names(mf), 0)
  mf <- mf[c(1,m)]
  mf$drop.unused.levels <- TRUE
  mf[[1]] <- quote(stats::model.frame)
  mf <- eval.parent(mf)

  mt <- attr(mf, "terms")
  X <- model.matrix(mt, mf, contrasts)[,-1]
  Y <- model.response(mf, 'numeric')

  if(is.null(id)){id<-paste(1:length(Y))}

  X<-as.matrix(cbind(X))
  p<-dim(X)[2]
  N<-length(unique(id))
  NM<-length(Y)
  s<-rank(Y)

  if(p==1){warning("Test only available for more than one covariate")
    break}

########################


  ## Using (p-1) dimensional polar coords
sn_pdim<-function(mythetavec, psi=rep(1,N), h_mult=1){

    h= h_mult*(N^(-1/3))
    reppsi<-c(unlist(apply(cbind(1:N),1,function(x) rep(psi[x],sum(id==unique(id)[x])))))
    betavec<-cbind(rev(c(polar2rect(1, mythetavec))))

    okaym<-function(m){ sum(reppsi*reppsi[m]*(s>s[m])*pnorm((X%*%betavec-c(X[m,]%*%betavec))/h))}
    bb<-sum(-apply(cbind(1:NM),1,okaym))/((N)*(N-1))
    return(bb)
  }
  ##########################
  
 generateoriginal<-function(my_H){
  
  sn_pdim_H<-function(x){sn_pdim(x, psi=rep(1,N), h_mult=my_H)}
  
  init_points<-matrix(runif(6*(p-1), -3.142,  3.142),6,)
  trial_outs<-init_points*0
  trials<-rep(0,6)
  for(jj in 1:dim(init_points)[1]){
  	
  	tryouts<-nlm(sn_pdim_H, init_points[jj,], iterlim=3)
  	trial_outs[jj,]<-tryouts$estimate
  	trials[jj]<-tryouts$minimum
  	
  }
  theta_hat = nlm(sn_pdim_H, trial_outs[which.min(trials),])$estimate
  mybetavec<-rev(polar2rect(1,theta_hat))
  yfit = c(X%*%mybetavec)
  return(list(sd=sd(yfit), point_est=mybetavec))}
  
  ##########################

myH<-generateoriginal(1)$sd
print("Part 1: establishing the bandwidth and point estimate.")
for(q in 1:Q){
part1_out<-generateoriginal(myH)
myH_new<-part1_out$sd
myH<-myH_new
print(paste("sigma[",q,"]=", round(myH,3), sep=""))}
my_betavec<-part1_out$point_est
print(paste("point estimate ="))
print(round(my_betavec,4))

print("Part 2: Calculating p-values.")
  mybetavec_b<-matrix(0,B,p)
  pb = txtProgressBar(min = 0, max = B, initial = 0)
  for(bb in 1:B){
    setTxtProgressBar(pb,bb)

    tryCatch({
      psib=rexp(N)
      snB_pdim<-function(x){sn_pdim(mythetavec=x, psi=psib, h_mult= myH)}

        init_points<-matrix(runif(6*(p-1), -3.142,  3.142),6,)
  trial_outs<-init_points*0
  trials<-rep(0,6)
  for(jj in 1:dim(init_points)[1]){
  	
  	tryouts<-nlm(snB_pdim, init_points[jj,], iterlim=3)
  	trial_outs[jj,]<-tryouts$estimate
  	trials[jj]<-tryouts$minimum
  	
  }
      theta_hat_b = nlm(snB_pdim, trial_outs[which.min(trials),])$estimate
      mybetavec_b[bb,]<-rev(polar2rect(1, theta_hat_b))


    }, error=function(e){cat("ERROR :",conditionMessage(e), "\n")})
  }
  close(pb)

  pval1 <-rep(0,p)
  for(pp in 1:dim(X)[2]){
    pval1[pp]<-	2*min(c(   (1+ sum(mybetavec_b[,pp]>0))/(1+B) ,  (1+ sum(mybetavec_b[,pp]<=0))/(1+B)  ))
  }

  CI<-apply(mybetavec_b,2,function(x) quantile(x,c(0.025,0.975)))
  result<-data.frame(scaled_estimate=c(my_betavec), lower95CI=CI[1,],upper95CI=CI[2,], pval= pval1)
  rownames(result)<-colnames(X)

  print(paste("Number of Observations:",NM))
  print(paste("Number of Groups:",N))

  return(result)}

####################################################
		
	\end{verbatim}	
	\end{footnotesize}

\end{document}